\documentclass[prd,aps,showpacs,tightenlines,nofootinbib]{revtex4}
\usepackage{graphicx}

\newcommand{\bear}{\begin{array}}  \newcommand{\eear}{\end{array}}
\newcommand{\bea}{\begin{eqnarray}}  \newcommand{\eea}{\end{eqnarray}}
\newcommand{\beq}{\begin{equation}}  \newcommand{\eeq}{\end{equation}}
\newcommand{\bef}{\begin{figure}}  \newcommand{\eef}{\end{figure}}
\newcommand{\bec}{\begin{center}}  \newcommand{\eec}{\end{center}}
\newcommand{\non}{\nonumber}

\newcommand{\del}{\partial}  

\newcommand{\bib}{\bibitem}

\def\IBIDD#1#2#3{{\it ibid}. {\bf #1}, #2 (20#3)}

\def\APJ#1#2#3{Astrophys. J. {\bf #1}, #2 (19#3)}

\def\APJLL#1#2#3{Astrophys. J. Lett. {\bf #1}, L#2 (20#3)}

\def\NATT#1#2#3{Nature (London) {\bf #1}, #2 (20#3)}
\def\NPB#1#2#3{Nucl. Phys. {\bf B#1}, #2 (19#3)}

\def\PLB#1#2#3{Phys. Lett. B {\bf #1}, #2 (19#3)}

\def\PRD#1#2#3{Phys. Rev. D {\bf #1}, #2 (19#3)}
\def\PRDD#1#2#3{Phys. Rev. D {\bf #1}, #2 (20#3)}

\def\PRLL#1#2#3{Phys. Rev. Lett. {\bf#1}, #2 (20#3)}

\newcommand{\ds}{\displaystyle}

\begin{document}

\title{I-balls}
\author{S. Kasuya$^a$, M. Kawasaki$^b$ and Fuminobu Takahashi$^b$}
\affiliation{$^a$ Helsinki Institute of Physics, P.O. Box 64, FIN-00014,
University of Helsinki, Finland\\
$^b$ Research Center for the Early Universe, University of Tokyo,
Tokyo 113-0033, Japan}
\date{\today}

\begin{abstract}
    We find that there exists a soliton-like solution ``I-ball''
    in theories of a real scalar field if the scalar potential satisfies
    appropriate conditions.  Although the I-ball does not have any
    topological or global $U(1)$ charges, its stability is ensured by
    the adiabatic invariance for the oscillating field.
\end{abstract}

\maketitle


\section{Introduction}
\label{sec:introduction}

Scalar fields play important roles in theories of the early universe. It
is believed that our universe experienced quasi-exponential expansion
phase ($=$ inflation ) in its very early stage, which solves the
flatness and horizon problems of the standard cosmology and explains the
origin of the density fluctuations of the universe such as observed by
COBE~\cite{COBE} and other experiments~\cite{BOOMERANG,MAXIMA,DASI}. The
inflationary universe scenario is realized by the vacuum energy of some
scalar field (inflaton). After inflation, the inflaton starts to
oscillate and decays into other particles which reheat the universe
through thermalization processes.

Similar dynamics is found in the Affleck-Dine mechanism for
baryogenesis~\cite{AD} which is a promising candidate for explaining the
matter-antimatter asymmetry of the universe. The mechanism makes use of
a scalar field (AD field) corresponding to a flat direction in the scalar
potential of the minimal supersymmetric standard model. During inflation
the AD field has a large field value and oscillates when the effective
mass becomes smaller than the Hubble parameter after
inflation. When the AD field starts oscillation, the baryon number is
generated through the baryon number violating term in the potential. 

Recently, it was found that the oscillating AD field deforms into lumps
of the scalar condensate called Q
balls~\cite{Kusenko,Enqvist,Kasuya1,Kasuya2}. The Q ball is a
non-topological soliton and its stability comes from the global charge
($=$baryon number) conservation. The existence of the Q ball is crucial
because it may significantly change the scenario of the Affleck-Dine
baryogenesis~\cite{Kasuya3}. The fragmentation into scalar lumps may
also take place for the inflaton field. In fact, Enqvist {\it et
al}~\cite{EnqvistKasuya} pointed out that the oscillating inflaton field
can fragment into Q balls.

Since the Q ball is stable owing to the charge conservation, the scalar
field responsible for the Q ball must be complex. Then, a question arises
whether or not a real scalar field deforms into lumps similar to the Q
balls. At first glance, no stable lumps are formed because any
conservation quantities like a global charge do not exist for the system
of a real scalar field.  However, the previous studies on the dynamics of
scalar fields showed that some soliton-like objects are formed. For
example, ``oscillons'' are formed for 
phase the double well potentials~\cite{Gleiser} and the axion field
fragments into ``axitons''~\cite{Tkachev}. In both cases the numerical
simulations showed the existence of some scalar lumps inside which the
scalar fields are rapidly oscillating. However, the reason why such
quasi-stable scalar lumps are formed was not clear at all. Moreover,
recently, McDonald~\cite{McDonald} pointed out that in a hybrid inflation
model the inflaton field can fragment into scalar lumps even if the scalar
field has any conserved charges [see also Ref.~\cite{rajantie}].

Thus, it has been seen that real scalar fields fragment into quasi-stable
lumps in numerical simulations for various situations.  In this paper,
therefore, we face the important problem concerning the real scalar
dynamics, that is, what makes the scalar lump quasi-stable?  Because the
conserved baryon number plays a crucial role for stability of the Q ball,
we need similar conservation quantity to stabilize the real scalar
lump. In classical mechanics it is well-known that the adiabatic invariant
exists for oscillating phenomena~\cite{landau}.  As will be seen later, we
find that the adiabatic invariant can be extended to the field
theories. Then, the existence of the stable lump can be explained by the
adiabatic charge $I$ ( see the next section for definition ) for the
oscillating scalar field.  We call this scalar lump ``I-ball'', since the
adiabatic charge $I$ plays the same role as the global $U(1)$ charges in
the case of Q balls.  We obtain the condition on the form of the scalar
potential for the I-ball formation and derive the equation which
determines the field configuration of the I-ball.  In particular, it is
found that the adibaticity requires the scalar potential to be dominated
by a quadratic term.  We also perform numerical simulations to confirm the
existence of the I-balls for two types of simple potentials.

\section{Adiabatic Charge}
\label{sec:adiabatic}

In this section we derive the conservation of an adiabatic charge, which
guarantees the stability of the I-ball as discussed later.  First we
shortly review an adiabatic invariant in a mechanical system, according to
Ref.\cite{landau}. Let us suppose that a system is executing a finite
motion in one dimension and characterized by some parameter $\lambda(t)$
which specifies the properties of the external field. We assume that
$\lambda$ varies slowly enough ({\it i.e.},``adiabatically''):
\begin{equation}
\left|\frac{\dot{\lambda}}{\lambda}\right| \ll T^{-1},
\end{equation}
where an overdot represents a derivative with respect to time, and $T$ is
the period of the motion.  If $\lambda$ is constant, the system executes a
strictly periodic motion with a constant energy $E$.  For slowly varying
$\lambda$, the energy $E$ varies slowly, while there is a quantity
which remains constant, called an adiabatic invariant. This is written as
\begin{equation}
I_{\rm 1dim} \equiv \frac{1}{2 \pi}\oint p ~dq,
\end{equation}
where $q$ and $p$ are the coordinate and momentum, and the integral is
taken over the variation of the coordinate during one period.  Note that
the periodicity plays an essential role in the proof of the existence for
the adiabatic invariant. Before we go on to the case of the field theory,
it will be useful to discuss a multi-dimensional system. Let us consider a
system with any number of degrees of freedom $\{q_i,p_i\}$, executing a
finite motion in all the coordinates.  It is assumed the variables are
separable so that the action can be written as the sum of functions each
depending on only one coordinate. As shown in Ref.~\cite{landau}, the
motion of the system is in general not strictly periodic, but {\it
conditionally periodic} since the system passes arbitrarily close to a
given state in the corse of a sufficient time.  In fact, it is periodic
only if the frequencies of all degrees of freedom are commensurable for
arbitrary values of $\{q_i\}$. In this case, there exists only one
adiabatic invariant in the system. On the other hand, if the variables are
not separable, there is no adiabatic invariant in general. However, if the
Hamiltonian of the system differs only by small terms from one which
allows separation of the variables, the properties of the motion are close
to the periodic motion.

Thus, in order to extend the adiabatic invariant to the case of the field
theory, we have to impose the following two assumptions. First, the
gradient energy is always sub-dominant everywhere, since the action of a
scalar field is separable except the gradient term.  Second, the scalar
potential is quadratic, which would ensure the strictly periodic motion if
it were not for the gradient term.  In the following arguments, we adopt
these two assumptions.

Let us consider the system of a real scalar field $\phi$, whose motion is
finite and characterized by a parameter $\lambda(\mathbf{x},t)$. We
suppose that $\lambda$ varies adiabatically, and that its dependence on
the position ${\bf x}$ is weak enough as well.  In the limit of constant
$\lambda$, the motion of the scalar field is homogeneous and periodic with
the period $T$.  The Lagrangian is given as,
\begin{eqnarray}
{\mathcal L } &=& \frac{1}{2} \partial_\mu \phi \partial^\mu \phi-V(\phi,\lambda),\\
\left.V(\phi,\lambda)\right|_{t<0} &=& \frac{1}{2} m^2 \phi^2\,,
\end{eqnarray}
where we assumed that the potential is quadratic with the mass equal to
$m$ before the external field $\lambda(\mathbf{x},t)$ is turned on at
$t=0$. If $\lambda$ were constant, the energy-momentum conservation law
would be given as
\begin{eqnarray}
 \partial_{\nu} T^{\mu \nu} &=& 0,\\
 T^{\mu \nu}& \equiv& \partial^\mu \phi \, \partial^\nu \phi - \eta^{\mu \nu} {\mathcal L }.
\end{eqnarray}
We turn attention to its time component, 
\begin{eqnarray}
 \partial_{\mu} j^{\mu} &=& 0,\\
 j^{\mu}& \equiv& T^{\mu 0} = T^{0 \mu} =
 \dot{\phi} \, \partial^\mu \phi - \eta^{\mu 0} {\mathcal L },
\end{eqnarray}
where an overdot represents $\partial_t$, and $\eta_{\mu \nu} = {\rm
diag}(+,-,-,-)$. With slowly varying $\lambda$ for $t > 0$, the motion of
the scalar field changes adiabatically. Then the energy-momentum current
$j^\mu$ is no longer conserved:
\begin{equation}
\label{eq:nocon}
\partial_{\mu} j^{\mu} = 
(\partial_\mu \lambda) \frac{\partial j^\mu}{\partial \lambda}.
\end{equation}
If we average Eq. (\ref{eq:nocon}) over the period $T$, we have
\begin{equation}
\label{eq:time}
\overline{\partial_{\mu} j^{\mu}} = 
\ds{
(\partial_\mu \lambda) \frac{1}{T}
\int_t^{t+T} dt \, \frac{\partial j^\mu}{\partial \lambda}
}
=\ds{
(\partial_\mu \lambda) 
\left(\oint \frac{d \phi}{\dot{\phi}}\right)^{-1}
\oint \frac{\partial j^\mu}{\partial \lambda} \frac{d \phi}{\dot{\phi}}}
,
\end{equation}
where the overline represents the average over the period of the motion:
\begin{equation}
\overline{Z} \equiv \frac{1}{T} \int_t^{t+T} dt \,Z.
\end{equation}
In deriving Eq. (\ref{eq:time}), we used the two assumptions, one of which
states that $\lambda$ varies adiabatically:
\begin{equation}
\label{eq:adiabatic-time}
\left|\frac{\dot{\lambda}}{\lambda} \right| \ll T^{-1}.
\end{equation}
Therefore, $\lambda$ can be regarded as a constant, when $\oint d \phi$ is
integrated over the total motion of $\phi$ during one period at fixed
${\mathbf x}$. The other assumption is that $\lambda$ depends on the
position weakly enough:
\begin{equation}
\label{eq:adiabatic-space}
 \left| \frac{\nabla \lambda }{\lambda}\right| \ll T^{-1},
\end{equation}
where $T$ can be interpreted as a typical spatial scale of the
oscillating system. This condition is necessary because otherwise the
gradient term becomes too large to be negligible. Note that the large
gradient term also changes the value of $\overline{\phi^2}$ significantly
during one period, leading to the violation of the adiabaticity.  For a
small but nonzero $\nabla \lambda$, the motion is not strictly periodic,
and the deviation from the orbit obtained as if $\lambda$ were constant
can be estimated to be of the order of $\delta \phi \sim \phi T \nabla
\lambda$ after one period. We neglect such small corrections, and would
like to focus attention on the leading terms in the following argument.
Thus the motion of $\phi$ is approximated to be both periodic over $T$ and
homogeneous in the volume $V=[{\mathbf x-T/2},{\mathbf x+T/2}]$. Hence
there are four conserved quantities $J^{\mu} \equiv \int_{V} T^{\mu 0}
d^{3}x$ along the path of the motion, and one can regard $\del^\mu \phi$
as a function of $(\phi,J^\mu,\lambda)$.

By differentiating the equation :
$j^\nu(\phi,\partial_\mu \phi, \lambda)  = J^\nu$ with $\lambda$,
we have
\begin{eqnarray}
   &&\frac{\partial j^\nu}{\partial \partial_\mu \phi} 
   \frac{\partial\partial_\mu \phi}{\partial \lambda}+ 
   \frac{\partial j^\nu}{\partial\lambda}=0,\non\\ 
   \Longleftrightarrow && \frac{\partial j^\mu}{\partial \lambda}
   =-\dot{\phi} \frac{\partial \partial^\mu\phi}{\partial \lambda}.
   \label{eq:diff}
\end{eqnarray}
In deriving  Eq. (\ref{eq:diff}), we have used the relation
\begin{equation}
   \frac{\partial  j^\nu }{\partial  (\partial^\mu \phi) }
   \simeq  \dot{\phi}  \,\delta^{\nu}{}_{\mu}.
\end{equation}
where the non-diagonal components are higher order in $\dot{\lambda}$ and
$\nabla \lambda$, and hence we neglect them.  With the use of
Eq. (\ref{eq:diff}), Eq. (\ref{eq:time}) can be rewritten as
\begin{eqnarray}
   \overline{\partial_{\mu} j^{\mu}} & =& 
   -\partial_\mu \lambda 
   \left(\oint \frac{d \phi}{\dot{\phi}} \right)^{-1}
   \oint \frac{\del (\partial^\mu \phi)}{\partial \lambda} d \phi
   \non\\
   \Longleftrightarrow &&
   \oint \left(
   \partial_\mu \lambda 
   \frac{\partial(\partial^\mu \phi)}{\partial \lambda} +
   \overline{\partial_{\mu} j^{\mu}}
   \frac{1}{\dot{\phi}} \right) d \phi =0\non\\
   \Longleftrightarrow &&
   \oint \left(
   \partial_\mu \lambda 
   \frac{\partial(\partial^\mu \phi)}{\partial \lambda} +
   \overline{\partial_{\mu} j^{\nu}}
   \frac{\partial (\partial^\mu \phi)}{\partial j^\nu}
   \right) d \phi =0.
\end{eqnarray}
Hence we have 
\begin{equation}
   \overline{\partial_\mu \left(
   \oint d\phi \,\partial^\mu \phi \right)} =0,
\end{equation}
where the time component ($\mu = 0$) is the leading term, and has a
definite physical meaning.  Thus, we are led to define the adiabatic
charge $I$ as
\begin{eqnarray}
   I &\equiv& \frac{1}{2m}\int d^3 x\,
~\overline{\dot{\phi}^2}, 
\end{eqnarray}
where the factor $1/m$ in the definition here is introduced to make $I$
dimensionless. Apparently $I$ is conserved,
\begin{equation}
   \frac{d I}{d t}=0\,. 
\end{equation}

\section{Conditions for existence of I-balls}
\label{sec:condition}

With the use of the adiabatic charge derived in the previous section, let
us consider the condition that I-balls are formed. We would like to focus
attention on $\lambda$ in the first place. It specifies the properties of
the external field, hence the energy of the system is no longer conserved
for varying $\lambda$ (see Eq.~(\ref{eq:nocon})). Alternatively, it could
be argued that the role of $\lambda$ is played by the self-interaction of
the scalar field, which must be such that the adiabatic conditions,
(\ref{eq:adiabatic-time}) and (\ref{eq:adiabatic-space}) are
satisfied. Then the total energy of the system including the interaction
would be conserved, although the energy of the free part varies due to the
self-coupling.  In other words, there are two invariants, the energy and
adiabatic charge for the whole system including `the external field'.

We assume that the scalar potential $V(\phi)$ is given as
\begin{equation}
V(\phi)=\frac{1}{2} m^2 \phi^2 + V_1(\phi),
\end{equation}
where the self-interaction $V_1(\phi)$ is small enough to respect the
adiabatic conditions.  That is to say, the adiabatic conditions are
assumed to be satisfied by the dynamics of the system.  First, we separate
$\phi$ into the rapidly oscillating part ($\tilde{P}$) and slowly varying
part ($\Phi$) as
\begin{equation}
\phi({\mathbf x},t)=\Phi({\mathbf x},t)\tilde{P}({\mathbf x},t).
\end{equation}
For the general potentials, $\tilde{P}$ might strongly depend on ${\mathbf
x}$, which leads to the large gradient energy $(\nabla \phi)^2$ and
violates the adiabatic condition (\ref{eq:adiabatic-space}).
However, we can safely adopt the ansatz that $\tilde{P}$ is homogeneous
over a sufficiently large scale in which we are interested, since the
periodicity of the system is guaranteed if the adiabatic conditions are
satisfied. Thus,
\begin{equation}
  \tilde{P}({\mathbf x},t) = P(t) 
\end{equation}
where $P(t)$ oscillates between $-1$ and $1$ with the period $T$.  We also
define O(1) constants $c_n$ ( $n = 1, 2, 3, \cdots$ ) for
the later use.
\begin{eqnarray}
\overline{\phi^{2n}}  & =& c_n \Phi^{2n},\\
\overline{\dot{\phi}^2}  & =& c_1 m^2 \Phi^2,
\end{eqnarray}
where we used the fact that the scalar potential is dominated by
the quadratic term.

\begin{figure}
\includegraphics[width=10cm]{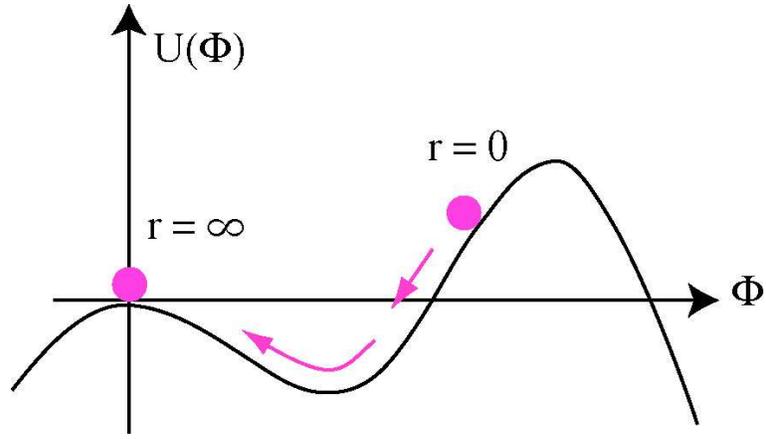}
\caption{\label{fig:potential} 
Potential $U(\Phi)$}
\end{figure}

We take advantage of the method of Lagrange multipliers to look for the
minimum of the energy $E$ at fixed $I$, and minimize
\begin{eqnarray}
   E_\omega &=& \overline{E} 
   +
   \tilde{\omega} \left(I -  \frac{1}{2m}\int d^3 x \, \overline{\dot{\phi^2}}
   \right)\,,\non\\
   &=& \int d^3 x \left(
   \left(1-\frac{\tilde{\omega}}{m}\right) \frac{1}{2}\overline{\dot{\phi^2}}
   +\frac{1}{2}  \overline{|\nabla \phi|^2}
   + \overline{ V(\phi)}
   \right)+
   \tilde{\omega}I\,,\non\\
   &=& \int d^3 x \,c_1 \left(
   \frac{1}{2}   |\nabla \Phi|^2 
    -\frac{\omega m}{2} \Phi^2 
   + V(\Phi)\right)
   +\left(\omega+m\right) I\,,
\end{eqnarray}
where $\tilde{\omega}\equiv \omega+m$, and $V(\Phi) \equiv \overline{
V(\phi)} /c_1 $.  Assuming the bounce solution is spherically symmetric,
\begin{equation}
  E_\omega = 
  \int dr 4 \pi r^2 \, c_1 \left(
  \frac{1}{2}  \left( \frac{d\Phi}{dr}  \right)^2
  +V(\Phi)
-\frac{\omega m}{2} \,  \Phi^2 
\right)+\left(\omega + m\right) I\,.
\end{equation}
In order to minimize $E_\omega$ with respect to $\Phi$, we have to seek
for the bounce solution, which is equivalent to solving the equation,
\begin{equation}
   \label{eq:eq-of-mot}
   \frac{d^2 \Phi}{d\, r^2}+\frac{2}{r} \frac{d\Phi}{dr}
   +\frac{dU}{d\Phi}=0\,,
\end{equation}
where
\begin{eqnarray}
   \label{eq:potential}
	U(\Phi) &\equiv& \frac{\omega m}{2}\, \Phi^2
   -V(\Phi)\,.
\end{eqnarray}
The bounce solution should satisfy the following boundary conditions [see
Fig.~\ref{fig:potential}]:
\begin{eqnarray}
   \left. \frac{d \Phi}{dr}\right|_{r=0}&=&0\,,\\
   \Phi(\infty)&=&0\,.
\end{eqnarray}
Hence, the bounce solution (I-ball solution) exists if the following
inequalities are satisfied.
\begin{equation}
  \label{eq:ibc} 
   {\rm min}\left[\frac{2 V(\Phi)}{\Phi^2}\right] < \omega m
   < m^2 .
\end{equation}

Now we investigate the constraint on the interaction $V_1$.
For the field configuration to satisfy the adiabatic condition
(\ref{eq:adiabatic-space}), we require $|(d\Phi/dr)/\Phi| \ll m$.
Since the spatial scale of the I-ball solution is determined by
the mass scale of $U(\Phi)$, this requirement is equivalent to
\begin{equation}
\left|\frac{d^2 U(\Phi)}{d\,\,\Phi^2} \right| \sim 
\left|\frac{d^2 V_1(\Phi)}{d\,\,\Phi^2} \right| \ll m^2
\end{equation}
where we used the inequality (\ref{eq:ibc}). If the interaction
satisfies this constraint, the periodicity of the system is 
maintained, and the I-ball configuration which minimizes the total
energy of the system is also gently-sloping enough.

Lastly, we comment on the cases that the scalar potential is not dominated
by the quadratic term. Up to here we have shown that the adiabatic
invariant can be found for a scalar field with the quadratic potential in
the external field, and that the I-ball configuration minimizes the energy
of the whole system including the interaction, which plays a role of the
external field, for the fixed adiabatic invariant. Thus it is not evident
from the preceding arguments whether the lumps are formed for other
potentials. However, we performed the numerical calculations, and found
that no quasi-stable lumps are formed for the several examples of such
potentials. Hence this fact strongly suggests that the existence of
I-balls is peculiar to the case where the quadratic term dominates the
potential.

\section{Numerical Simulation}
\label{sec:num}

Now that we have shown that the I-balls minimize the energy for fixed $I$,
it must be then investigated whether such soliton-like objects are really
formed. For this purpose we perform numerical calculation which follow the
time evolution of the system. This is a quite nontrivial question, since
the requirement that the system should vary adiabatically and its spatial
distribution be gently-sloping enough ( ``adiabaticity condition'' ) must
be satisfied dynamically.  As concrete examples we take both the
gravity-mediation like potential and $m^2\phi^2-\phi^4$ potential.

\begin{figure}
\includegraphics[width=10cm]{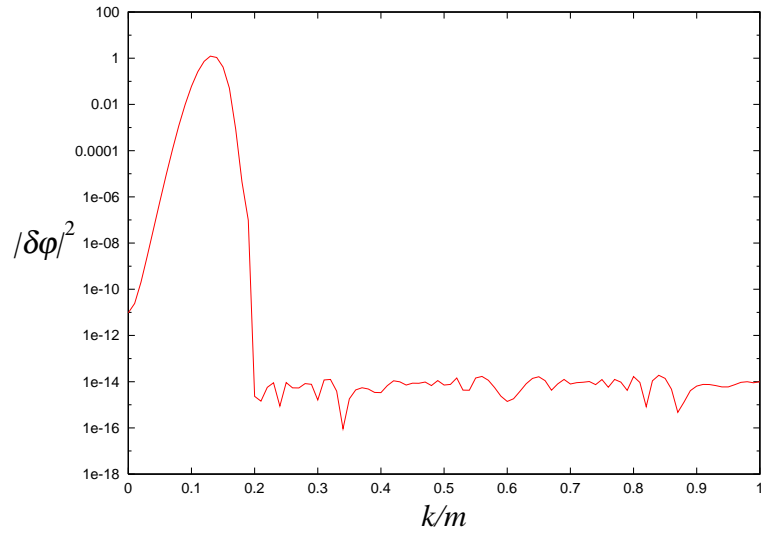}
\caption{\label{fig:gra-band} 
Instability band for the gravity-mediation like potential with $K=-0.1$.}
\end{figure}

First let us suppose the potential is given as
\begin{eqnarray}
V(\phi)&=&\frac{1}{2} m^2 \phi^2 \left(
1+K \log\left(\frac{\phi^2}{2M_*^2}\right)
\right)\,,
\label{eq:grapote}
\end{eqnarray}
where $M_*$ is a renormalization point to define the
mass, and the $K$ term is the one-loop correction which
is assumed to be negative.
The I-ball equation Eq.~(\ref{eq:eq-of-mot}) reads
\begin{equation}
\label{eq:gra-type-iballeq}
\frac{d^2 \Phi}{dr^2}+\frac{2}{r} \frac{d \Phi}{dr}
+\omega_0^2 \Phi-m^2 \Phi K \log\left(
\frac{\Phi^2}{4M_*^2}
\right)=0\,,
\end{equation}
where $\omega_0$ is defined as
\begin{equation}
\omega_0^2\equiv  \omega m -m^2(1+K).
\end{equation}
From numerical calculations, it is suggested that a Gaussian ansatz is a
reasonable approximation to the I-ball solution for this potential.  If we
insert the Gaussian ansatz,
\begin{equation}
\Phi(r)=\Phi(0) e^{-r^2/R_I^2}\,,
\end{equation}
into the I-ball equation Eq.~(\ref{eq:gra-type-iballeq}), we obtain
\begin{equation}
\left(\frac{4 r^2}{R_I^4}-\frac{6}{R_I^2}\right)\Phi
+\left(\omega_0^2+\frac{2K m^2}{R_I^2}r^2\right) \Phi=0\,,
\end{equation}
where we have set $M_*=\Phi(0)/2$.  Thus we see that the same form
is obtained in the first and second terms.  This requires that
\begin{eqnarray}
R_I &=& \frac{\sqrt{2}}{\sqrt{|K|}m}\,,
\end{eqnarray}
which roughly corresponds to the inverse of the most amplified mode
$k_{\rm max}$, and this correspondence is  checked numerically [see
Fig.~\ref{fig:gra-band}]. Also the maximum growth rate of the linear
perturbation is estimated to be of the order $|K|m$ in the similar way as
in the case of Q balls~\cite{Enqvist,EnqvistKasuya}.  Hence the
adiabaticity condition is satisfied for the gravity-mediation like
potential.

\begin{figure}
\includegraphics[width=10cm]{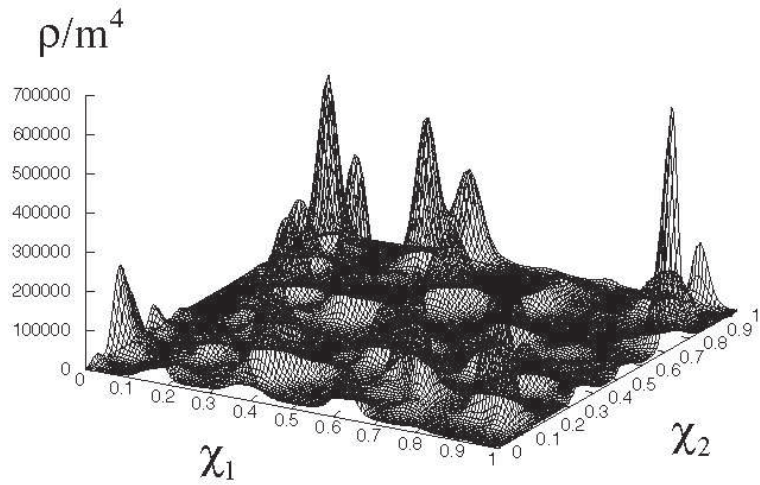} \caption{\label{fig:gra} Spatial
distribution of the energy density $\rho$ for the gravity-mediation like
potential. Inside the I-balls, the scalar field $\phi$ oscillates
rapidly.}
\end{figure}

\begin{figure}
\includegraphics[width=10cm]{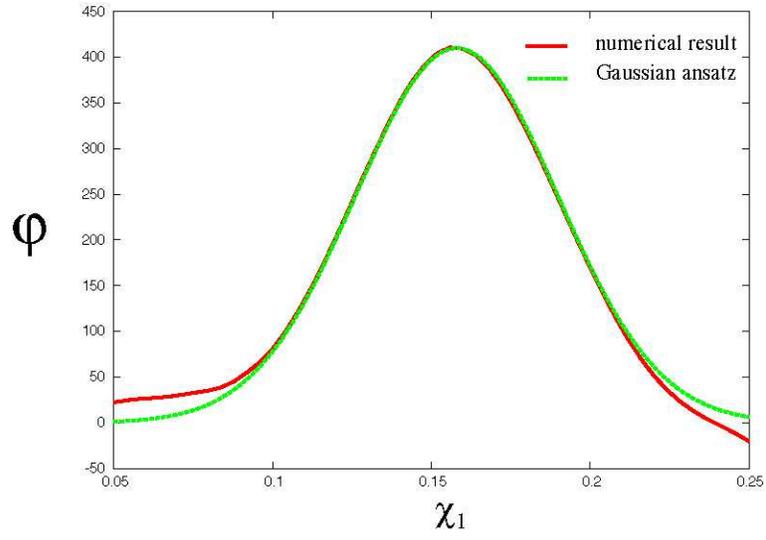} \caption{\label{fig:profile} Profile
of the scalar field inside the typical I-ball.  The analytic solution
(dotted line) agrees very well with the actual profile (solid line).  }
\end{figure}

We perform the numerical simulation to confirm whether the I-balls are
formed with the radius obtained above. For the numerical
calculation, we take the variables to be dimensionless as follows.
\begin{eqnarray}
\label{eq:norm}
\varphi &=& \frac{\phi}{m},\non\\
\tau &=& m t,\non\\
\chi_i &=& m x_i,\non\\
\kappa_i &=& k_i/m,
\end{eqnarray}
where $k_i$ is a wave number in the $x_i$ direction. The initial
conditions are taken as
\begin{eqnarray}
\varphi(0) & =& 1.6\times 10^5 + \delta_1,\non\\
\varphi'(0) &=& -\frac{1}{3} + \delta_2,
\end{eqnarray} 
where the prime denotes the derivative with $\tau$, and we set $K = -0.1$.
$\delta$'s are fluctuations which originate from the quantum fluctuations,
and their amplitudes are taken to be $10^{-7}$ times smaller than the
homogeneous mode. We have confirmed that the smaller fluctuations just
delay the formation of I-balls. First we check that the adiabatic
condition is satisfied up to the linear growth of
perturbations. Fig.~\ref{fig:gra-band} shows the numerical result of the
instability band for the initial conditions stated above.  It can be seen
that the instability band roughly corresponds to the inverse of the radius
$R_I$ obtained analytically.  We present the result of numerical
simulation in two dimensional lattices in Fig.~\ref{fig:gra}, from which
one can see that the energy density $\rho$ deforms into lumps, identified
as I-balls. Also Fig.~\ref{fig:profile} represents the profile of the
scalar field $\Phi$ inside the I-ball, and the analytic solution obtained
above agrees quite well with the numerical result, which suggests the
Gaussian ansatz is appropriate.
Though there is some deviation in the outer region, it is irrelevant since
the absolute value of the scalar field is much smaller than that at the
center of the I-ball. \footnote{Some reasons can be adduced: (1) The
adiabatic conditions might not be satisfied well around the surface, since
the homogeneous mode is relatively small compared to the fluctuations.
(2) Actually the I-balls are not isolated. }

\begin{figure}
\includegraphics[width=10cm]{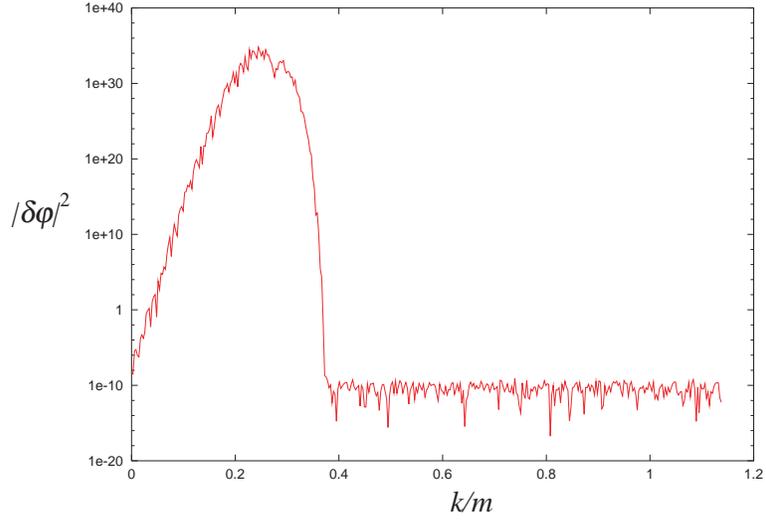}
 \caption{\label{fig:linear} Instability band for $\phi_i=m$,
 $\dot{\phi_i}=0$, $a=0.1$ and $b=0.005$ in the potential
 Eq.~(\ref{eq:pote}). }
\end{figure}

Next we take the following potential.
\begin{equation}
\label{eq:pote}
V(\phi)=\frac{1}{2}m^2 \phi^2 - \frac{a}{4} \phi^4 + \frac{b}{6 m^2}
 \phi^6\,, 
\end{equation}
where we have added the $\phi^6$ term for the stability of the vacuum.
Here $a$ is positive to satisfy the condition (\ref{eq:ibc}). Actually,
for negative $a$, no I-balls are produced in the numerical simulations.
As in the previous case, we take the variables to be dimensionless as
Eq.~(\ref{eq:norm}).  The initial conditions are taken as
\begin{eqnarray}
\varphi(0) & =& 1.0 + \delta_1,\non\\
\varphi'(0) &=& 0 + \delta_2.
\end{eqnarray} 
We set $a = 0.1$, $b = 0.005$, and $\delta$s' amplitudes are taken to be
$10^{-5}$ times smaller than the homogeneous mode. 
As long as $-\phi^4$ term is much smaller than the mass
term, the growth rate $\alpha$ and the instability mode ${\bf k}$ are
given by~\cite{McDonald}
\begin{eqnarray}
   \alpha({\bf k}) &\simeq& 
   \left( \frac{3 a \Phi^2}{8 m^2}\right)^{\frac{1}{2}} 
   |{\bf k}|,\non\\ 
   0<&{\bf k}^2&< \frac{3}{2} a \Phi^2\,.
\end{eqnarray}
Since we take $\phi = m$ as an initial condition, $\alpha({\bf k}_{\rm
max})$ and $\sqrt{{\bf k}_{\rm max}^2}$ are much smaller than $m$, so the
adiabaticity condition is satisfied up to linear perturbation.  The
numerical result of the instability band for the initial conditions stated
above is shown in Fig.~\ref{fig:linear}. We can see that the instability
band coincides exactly with that obtained analytically.

\begin{figure}
\includegraphics[width=10cm]{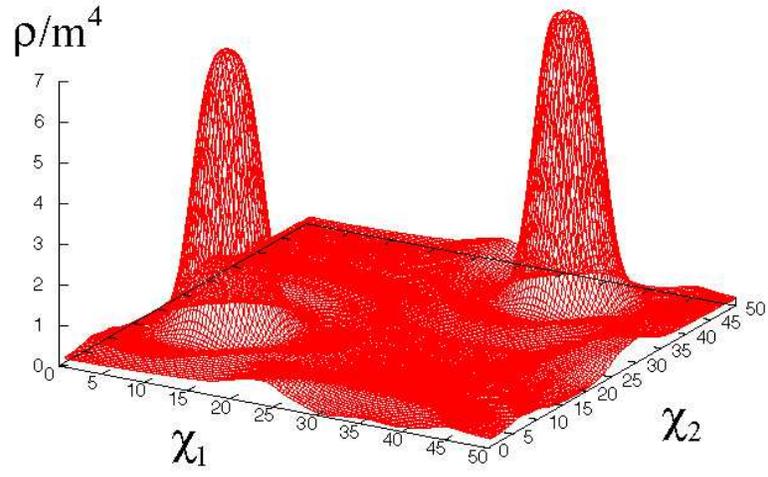}
 \caption{\label{fig:lattice} Spatial distribution of the energy density
 $\rho$ for the $m^2 \phi^2-\phi^4$ potential at time corresponding to
 $\tau = 450$, when the I-balls are newborn. Inside the I-balls, the
 scalar field $\phi$ oscillates rapidly.}
\end{figure}

We present the result of numerical calculation in two dimensional lattices
in Fig.~\ref{fig:lattice}. The energy density $\rho$ deforms into I-balls
in this case, too. We have performed numerical simulations with different
$a$, and found that all of these results generally look alike.
When the I-balls are newborn, they are almost spherically symmetric, and
the gradient energy is subdominant everywhere, suggesting that our
formulation is valid. As the system evolves in time, 
their shapes deviate from the spherically symmetric one, and become
irregular. Finally, they decay into random phases where the kinetic,
potential and gradient energies are all same order. The lifetime of the
I-balls $\tau$ becomes shorter and they are formed earlier as $a$
increases, but its typical value is $O(10^3 m^{-1})$ for $a=0.1$.  This is
consistent with the facts pointed out in Refs.~\cite{Gleiser,Tkachev},
that the oscillons and axitons have very long but finite lifetime. Hence,
it is certain that the I-balls for this type of potential decay in the
end.  The decay might be induced by the small deviation from the
adiabaticity of the dynamics, which induces the decoherence of the
oscillating scalar field inside I-balls.  But it needs further
investigations to make clear how the decay proceeds.

\section{Conclusion}
\label{sec:con}

We have studied the system of a real scalar field and found the solution
of the quasi-stable scalar lump, I-ball. The stability of the I-ball can
be explained by the adiabatic invariant charge $I$, which does stem from
the dynamics of the system, not any symmetries.  For the I-ball solution
to exist, the scalar potential should be dominated by the quadratic term
($m^2\phi^2$) and satisfy the condition (\ref{eq:ibc}) which is almost
same as that for the Q-ball.

Furthermore, we have performed numerical simulations
and have found that the quasi-stable I-balls are really produced and their
properties are in agreement with the theoretical predictions.  Since
scalar fields prevail in theories of the early universe, the I-balls may
be formed and play important roles in various cosmological
processes~\cite{KKT}.

\subsection*{ACKNOWLEDGMENTS}

M.K and F.T. thank Masahide Yamaguchi for useful discussion and
comments. Part of the numerical calculations was carried out on VPP5000 at
the Astronomical Data Analysis Center of the National Astronomical
Observatory, Japan.

\end{document}